\documentclass[aps,pra,superscriptaddress,twocolumn,showpacs,floatfix]{revtex4-2}

\usepackage{amssymb,amsmath,amsfonts,mathrsfs,bm, bbm, graphicx, epsfig, epstopdf}
\usepackage[version=3]{mhchem}
\usepackage{braket}
\usepackage{hyperref}
\usepackage{color}
\usepackage{xcolor}
\usepackage{adjustbox}
\usepackage{csquotes}
\usepackage{float}
\usepackage{colortbl}
\usepackage{siunitx}
\DeclareSIUnit{\nothing}{\relax}
\usepackage{makecell}
\usepackage[T1]{fontenc}
\usepackage{soul}

\date{\today}
\begin{document}
\title{One-Way Quantum Repeater with Rare-Earth-Ions Doped in Solids}

\author{Yisheng Lei}
\thanks{Corresponding author}
\email{YishengLei2025@u.northwestern.edu}
\affiliation{Department of Electrical and Computer Engineering and Applied Physics Program, Northwestern University, Evanston, IL 60208, USA}

\begin{abstract}
Quantum repeaters are proposed to overcome exponential photon loss over distance in fibers. One-way quantum repeaters eliminate the need for two-way classical communications, which can potentially outperform quantum memory based quantum repeaters. I propose that rare-earth-ions doped in solids and coupled with nano-cavity can be used to generate photonic cluster state efficiently, which serve as good platforms for one-way quantum repeater nodes. In addition, I propose a multiplexed scheme of photonic tree cluster state generation with multiple quantum emitters. With less than 100 quantum emitters, secret key rates can reach the order of MHz over a few thousand kilometers. This proposal is especially useful for generating large scale photonic cluster state, which is essential for correcting operational errors during processing in quantum repeater nodes.  
\end{abstract}

\maketitle{}

\section{Introduction}
Following the introduction of DLCZ, various types of quantum repeaters have been proposed \cite{briegel1998quantum, duan2001long}. Quantum repeaters can be classified into three generations: first generation: quantum memory based quantum repeaters with probablistic entanglement purification; second generation: quantum memory based quantum repeaters with deterministic entanglement purification; One-way quantum repeaters with deterministic error corrections \cite{muralidharan2016optimal}. Different schemes of one-way quantum repeater have been proposed over the past decade by using different physical systems: quantum parity code \cite{munro2012quantum, muralidharan2014ultrafast}, linear optics \cite{ewert2016ultrafast}, photonic graph state \cite{azuma2015all, buterakos2017deterministic, pant2017rate}, photonic tree-cluster state \cite{borregaard2020one, zhan2020deterministic}, GKP code \cite{rozpkedek2021quantum, fukui2021all} and hybrid state \cite{wo2023resource, niu2023all}. One-way quantum repeaters with photonic cluster states have attracted strong interests due to its promising future of implementations with current atomic systems. Multiple types of quantum emitters have been analyzed for generating photonic graph state, such as quantum dots, defect centers in diamond, cold atoms, etc. Rare-earth-ions doped in solids have wide applications in quantum information processing, such as ensemble-based quantum memories \cite{lei2023quantum}, quantum transduction \cite{bartholomew2020chip, rochman2023microwave}, single photon generation \cite{dibos2018atomic}, single-ion quantum registers \cite{kindem2020control, ruskuc2022nuclear}, and quantum computing \cite{kinos2021designing, kinos2022high}. In this article, I show that rare-earth-ions in solids can be excellent platforms for photonic cluster state generation. Furthermore, quantum repeaters with multiplexed quantum memory can greatly increase entanglement distribution rates and ease the requirement of long storage time \cite{collins2007multiplexed}, but one-way quantum repeater with multiplexed generation of error correction codes has never been analyzed, so I propose a multiplexed scheme of photonic cluster state generation. 

\section{Scheme}
Graph state (every vertex is a qubit with state $\mid$+$\rangle$ = $\frac{1}{\sqrt{2}}$($\mid$0$\rangle$ + $\mid$1$\rangle$, and every edge is a CZ gate between the two connected vertexes.) was initially introduced for one-way quantum computing \cite{raussendorf2001one} and loss-tolerance was further introduced \cite{varnava2006loss}. A number of photonic cluster generation schemes based on linear optics have been proposed and demonstrated experimentally \cite{browne2005resource, li2019experimental}. Due to its probabilistic processes, large overheads of experimental resources would be required \cite{pant2017rate}. Deterministic generation schemes with quantum emitters have been proposed recently \cite{lindner2009proposal, buterakos2017deterministic, borregaard2020one, zhan2020deterministic, vezvaee2022deterministic}, as well as hybrid approach \cite{hilaire2023near}. Various platforms have been demonstrated to generate photonic cluster states of small scales, such as cold atom \cite{thomas2022efficient} and quantum dot \cite{appel2022entangling, coste2023high, cogan2023deterministic}. Generation of large scale photonic cluster state requires the atomic system to have short excited state decay time, long spin coherence time and high quantum efficiency. Er$^\text{3+}$ ions in solids have atomic transitions with telecommunication C-band wavelength, which is ideal for photon transmissions in fibers. Er$^\text{3+}$ ions in CaWO$_\text{4}$ has been demonstrated with electron spin coherence time of 23 ms \cite{le2021twenty} and also when it is coupled with nanophotonic crystal cavity, indistinguishable single photons have been generated \cite{ourari2023indistinguishable}. Er$^\text{3+}$ ions doped in YSO crystal have been demonstrated with nuclear spin coherence time over one second \cite{ranvcic2018coherence}. Lifetime of Er$^\text{3+}$ ion excited state is $\sim$ 10 ms, so nano-cavity with large Purcell factor has to be depolyed to reduce it to be $\sim$ 1 ns in order to produce single photon in high rate. So far, Purcell enhancement factor of $\sim$ 1000 has been achieved with Er$^\text{3+}$ ion in MgO crystal \cite{horvath2023strong}.

\begin{figure}[!h]
\centerline{\includegraphics[width=0.95\columnwidth]{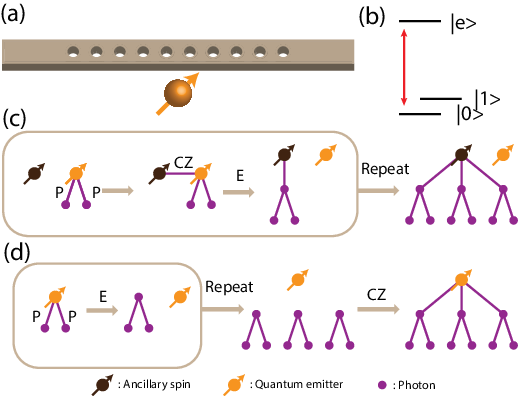}}
	\caption{(a) a single rare-earth ion coupled to a photonic crystal cavity. (b) atomic level structure of a $\Lambda$ system for photon-atom entanglement generation. (c) Photonic cluster state generation with one quantum emitter and one ancillary qubit. (d) Photonic cluster state generation with a single quantum emitter coupled into a cavity.}
	\label{Fig1}
\end{figure}

Er$^\text{3+}$ ion coupled with a nanophotonic crystal cavity is shown in Fig. \ref{Fig1}(a). A $\Lambda$ system with one excited state and two ground states (electronic spin splitting or nuclear spin splitting) is shown in Fig. \ref{Fig1}(b). Photonic time-bin qubits are more resilient to environmental noises compared with qubits of number states and polarization states. Here I describe the procedures of generating a photonic time-bin cluster state with a atomic $\Lambda$ system (note: Raman pulses can be used to control the transitions of the quantum emitter between states $\mid$0$\rangle$ and $\mid$1$\rangle$, or micro-electrodes can be integrated with the nanophotonic chips which can deliver rf pulses to induce transitions between the two ground states \cite{craiciu2021multifunctional}.): \\
Step 1: quantum emitter is initialized in state $\mid$0$\rangle$;\\
Step 2: a $\frac{\pi}{2}$ rf pulse is sent to the quantum emitter, $\frac{1}{\sqrt{2}}$($\mid$0$\rangle$ + $\mid$1$\rangle$);\\
Step 3: a $\pi$ optical pulse is sent to the quantum emitter, $\mid$e$\rangle$ or $\mid$1$\rangle$, and wait for decay;\\
Step 4: a $\pi$ rf pulse is sent to the quantum emitter, $\mid$1$\rangle$ if it is initially in $\mid$0$\rangle$, or $\mid$0$\rangle$ if it is initially in $\mid$1$\rangle$; \\
Step 5: a $\pi$ optical pulse is sent to the quantum emitter, $\mid$e$\rangle$ or $\mid$1$\rangle$, and wait for decay;\\
Step 6: a $\pi$ rf pulse is sent to the quantum emitter, $\mid$1$\rangle$ if it is initially in $\mid$0$\rangle$, or $\mid$0$\rangle$ if it is initially in $\mid$1$\rangle$; \\
Step 7: a $\pi$ rf pulse is sent to the quantum emitter, $\frac{1}{\sqrt{2}}$($\mid$0$\rangle$ + $\mid$1$\rangle$);\\

Single-qubit and two-qubit gate operations with rare-earth-ions in solids have been proposed \cite{kinos2021designing, kinos2022high}, and initial demonstrations with qubits of many rare-earth ions have been reported \cite{yan2021experimental, yan2019inverse}. Gate operation between photon and atom can be performed with atom in a cavity \cite{duan2004scalable}. Fig. \ref{Fig1}(c) shows generation scheme of a photonic tree cluster state with one quantum emitter and an ancillary spin \cite{buterakos2017deterministic}, and Fig. \ref{Fig1}(d) shows another scheme with only one quantum emitter coupled with a cavity \cite{zhan2020deterministic, zhan2023performance}. A photonic tree cluster state with tree depth of $d$ and branching parameters of ($n_0$, $n_1$, \dots , $n_{d-1}$), its generation time $T^2$ with the scheme in Fig. \ref{Fig1}(c) and $T^1$ with the scheme in Fig. \ref{Fig1}(d) \cite{zhan2023performance}: 
\begin{equation}
    T^2 \approx \prod_{i=0}^{d-1} n_it_P^2 + \left(\beta n_0 + \sum_{l=1}^{d-2} {\prod_{i=0}^l n_i}\right)t_E^2 + \sum_{l=0}^{d-2} {\prod_{i=0}^l n_i t_{CZ}^2},
\end{equation}

\begin{equation}
    T^1 \approx \prod_{i=0}^{d-1} n_it_P^1 + \sum_{l=1}^{d-2} {\prod_{i=0}^l n_i}t_E^1 + \sum_{l=0}^{d-2} {\prod_{i=0}^l n_i t_{CZ}^1},
\end{equation}

where $t_P$, $t_E$ and $t_{CZ}$ are the operation times of P gate (generate one photon entangled with the quantum emitter), E gate (generate one photon then perform state measurement to the quantum emitter to disentangle them) and CZ gate between photon and quantum emitter or between quantum emitter and ancillary spin. $\beta$ is the ratio between photons with longer wave packets to boost CZ gate fidelity and normal photons. Superscript 1 and 2 refer to the two schemes. 

\section{Multiplexing}
\begin{figure}[!h]
\centerline{\includegraphics[width=0.95\columnwidth]{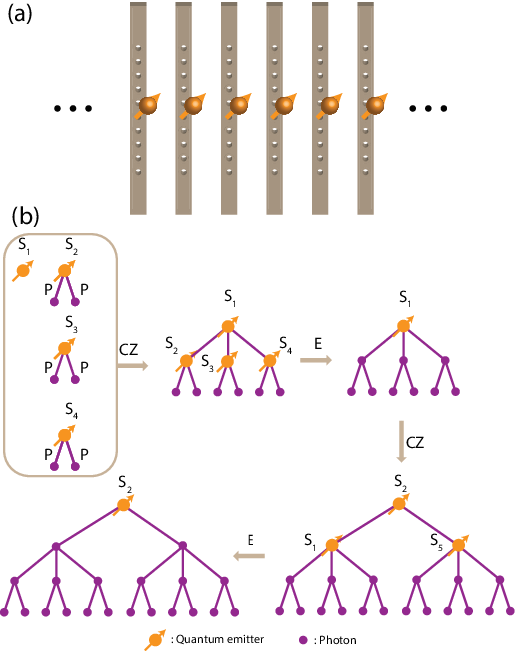}}
	\caption{(a) an array of single rare-earth ions coupled with nanophotonic crystal cavities. (b) generation procedures of a photonic tree cluster state with multiple quantum emitters.}
	\label{Fig2}
\end{figure}

The schemes based on single or a few atomic qubits would require long operation time for large photonic cluster state generation, which could cause strong decoherence errors limiting entanglement distribution rate and fidelity. Multiplexing with high number of quantum emitters has never been analyzed. Rare-earth ions have small footprints of $\sim$ 1 nm, and with the development of nano-cavity and integrated photonics, implantation of rare-earth ions in solids can be precisely controlled \cite{sullivan2023quasi, pak2022long}, multiplexing with a large number of quantum emitters become feasible. Here I propose a multiplexed scheme of photonic tree cluster state generation shown in Fig. \ref{Fig2} and analyze how multiplexing will affect secret key rate for one-way quantum repeaters.

Procedures with $\prod_{i=0}^{d-3} n_i(n_{d-2} + 1)$ quantum emitters are listed as following: \\
Step 1: starting from the bottom layer ($d$), each quantum emitter generate $n_{d-1}$ photons with P gate (photons in layer $d$ are all produced); \\
Step 2: at layer $d-1$, one quantum emitter performs CZ gate with each of the $n_{d-2}$ quantum emitters, each of which is attached with $n_{d-1}$ photons;\\
Step 3: after CZ gate, each of the $n_{d-2}$ quantum emitters performs E gate (photons at layer $d-1$ are all produced); \\
Step 4: at layer $d-2$, one quantum emitter performs CZ gate with each of the $n_{d-3}$ quantum emitters, each of which is attached with $n_{d-1}n_{d-2}$ photons;\\
Step 5: after CZ gate, each of the $n_{d-3}$ quantum emitters performs E gate (photons at layer $d-2$ are all produced); \\
Step 6: as we can see Step 4\&5 repeat Step 2\&3 for layer $d-2$. The two steps are repeated for each layer until layer $1$.\\
The total operation time:
\begin{equation}
    T^M \approx n_{d-1}t_P^M + \sum_{i=0}^{d-2} \left( n_{i}t_{CZ}^M + t_E^M \right),
\end{equation}
where M represents multiplexed scheme. From the equation, we can see that we only need to perform E gate for (d -1) times for a photonic tree cluster state of d-layer. This can be very useful, since state measurements usually take a much longer time compared with other gate operations. 

Procedures with $\prod_{i=0}^{d-3} n_i(\frac{n_{d-2}}{m} + 1)$ quantum emitters, for simplicity I assume $m$ and $\frac{n_{d-2}}{m}$ are integers: \\
Step 1: starting from the bottom layer ($d$), each quantum emitter generate $n_{d-1}$ photons with P gate (photons in layer $d$ are all produced); \\
Step 2: at layer $d-1$, one quantum emitter performs CZ gate with each of the $\frac{n_{d-2}}{m}$ quantum emitters, each of which is attached with $n_{d-1}$ photons;\\
Step 3: After CZ gate, each of the $\frac{n_{d-2}}{m}$ quantum emitters performs E gate; \\
Step 4: Repeat Step 1 - 3 for $m$ rounds (photons in layers $d$ and $d-1$ are all produced);\\
Step 5: at layer $d-2$, one quantum emitter performs CZ gate with each of the $n_{d-3}$ quantum emitters, each of which is attached with $n_{d-1}n_{d-2}$ photons;\\
Step 6: After CZ gate, each of the $n_{d-3}$ quantum emitters performs E gate (photons at layer $d-2$ are all produced); \\
Step 7: Repeat Step 5 - 6 for each layer until layer $1$.\\
The total operation time:
\begin{equation}
    T_m^M \approx m\left(n_{d-1}t_P^M + \frac{n_{d-2}}{m}t_{CZ}^M + t_E^M\right) + \sum_{i=0}^{d-3} \left( n_{i}t_{CZ}^M + t_E^M \right),
\end{equation}
where $m$ stands for the case which m rounds of photon generations for the bottom layer are performed.

\section{Repeater Performance}
For a photonic tree cluster state with tree depth of $d$ and branching parameters of ($n_0$, $n_1$, \dots , $n_{d-1}$), its transmission probability from one node to the following node is given \cite{varnava2006loss}: 
\begin{equation}
    \eta_t = \left[ \left(1 - \mu + \mu R_1 \right)^{n_0} - \left(\mu R_1 \right)^{n_0} \right] \left(1 - \mu + \mu R_2 \right)^{n_1}, 
\end{equation}
where $R_i$ is defined as the success probability of implementing an indirect Z measurement on any qubit in the i-layer. 
\begin{equation}
    R_k = 1 - \left[1 - \left( 1 - \mu \right) \left( 1 - \mu + \mu R_{k+2} \right)^{n_{k+1}} \right]^{n_k}, 
\end{equation}

with $R_{d+1} = 0$, $n_{d+1} = 0$ and $\mu = 1 - \eta_0\eta_s$, $\eta_0 = exp(-L_0/L_{att})$ is the single photon transmission probability between neighboring repeater node, and $\eta_s$ is the system quantum efficiency:  
\begin{equation}
    \eta_s = \eta_c \times \eta_w \times \eta_f  \times \eta_d, 
\end{equation}
where $\eta_c$ is the photon coupling efficiency to cavity mode, $\eta_w$ is the photon coupling efficiency from cavity to waveguide outside the cavity, $\eta_f$ is the photon coupling efficiency from waveguide to fiber and $\eta_d$ is the detection efficiency.

In this article, I consider quantum key distribution with six-state protocol \cite{bruss1998optimal, scarani2009security}. Its asymptotic key fraction is given by

\begin{equation}
    f = max \left\{ \left(1 - Q \right) \left[1 - h \left(\frac{1 - 3Q/2}{1 - Q} \right) \right] - h \left( Q \right), 0 \right\},
\end{equation}

where $h(x) = -x \text{log}_2 x - (1 - x)\text{log}_2(1-x)$ is the entropy function.
\begin{equation}
    \Delta_{\epsilon_p}(\rho) = (1 - \epsilon_p)\rho + \frac{\epsilon_p}{3}(\sigma_X\rho \sigma_X + \sigma_Y\rho \sigma_Y + \sigma_Z\rho \sigma_Z, 
\end{equation}
where $\Delta_{\epsilon}(\rho)$ is the photon error operator, $\rho$ is the photon density matrix and $\epsilon_p$ is the photon error probability. For a repeater with $N_{node}$ repeater stations, it is estimated that $\epsilon_p \approx (N_{node} + 1)\epsilon_r$, where $\epsilon_r$ is the photon error probability at each repeater station \cite{borregaard2020one, azuma2015all}. 
\begin{equation}
    Q = \frac{2\epsilon_p}{3}
\end{equation}

\begin{center}
\begin{tabular}{ c c c }
\hline\hline
 Quantity & Symbol & Value \\ 
 \hline
 Single Photon emission time & $t_p$ & 1 ns \\
 \hline
 E gate time & $t_E$ &  10ns \\ 
 \hline
 CZ gate time & $t_{CZ}$ & 10 ns \\
 \hline
 Spin coherence time & $T_2$ & 1s \\
 \hline
 Detector efficiency & $\eta_d$ & 0.98 \\
 \hline
 Photon cavity coupling efficiency & $\eta_c$ & 1\\
 \hline
 Cavity waveguide coupling efficiency & $\eta_w$ & 0.99\\
 \hline
 Waveguide fiber coupling efficiency & $\eta_f$ & 0.99\\
 \hline
 Fiber attenuation constant & $L_{att}$ & 20 km \\
 \hline
 Photon error probability & $\epsilon_{r}$ & 1 $\times 10^{-5}$ \\
 \hline
\end{tabular}
\end{center}

\begin{figure}[!h]
\centerline{\includegraphics[width=0.95\columnwidth]{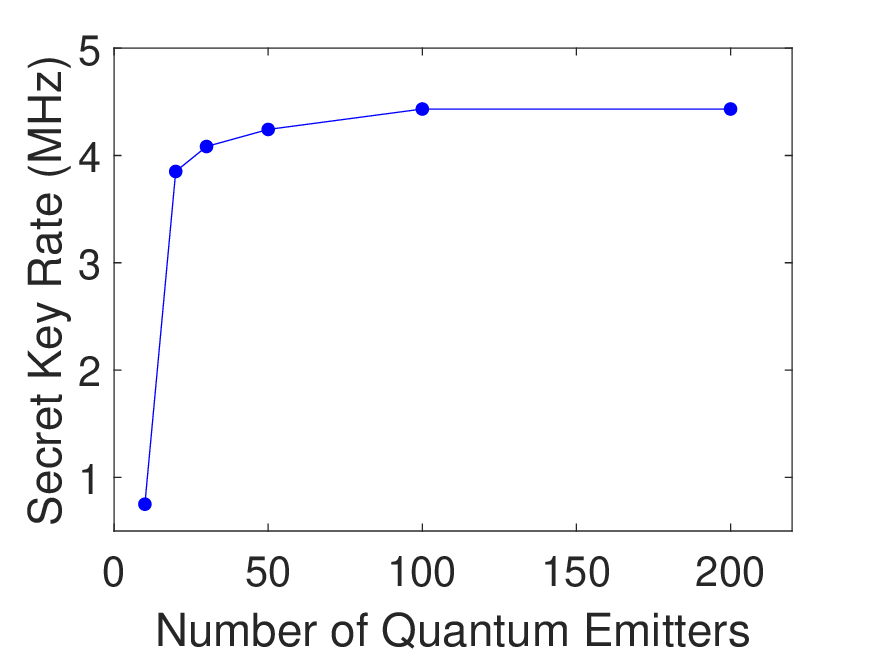}}
	\caption{(a) secret key rates with different numbers of quantum emitters for communication distance of 1000km.}
	\label{Fig3}
\end{figure}

Secret key rate is given by
\begin{equation}
    R = \frac{1}{T^M}f \eta_t^{N_{node} + 1},
\end{equation}
In Fig. \ref{Fig3}, the communication distance is fixed to be 1000km. Separations between neighboring repeater nodes are restricted to be above 1km.  I estimated the maximum secret key rates with number of quantum emitters between 10 and 200. The rate saturates after quantum emitters reach 20. This is due to the fast decay rate and small photon error probability in the simulation. 

\begin{figure}[!h]
\centerline{\includegraphics[width=0.95\columnwidth]{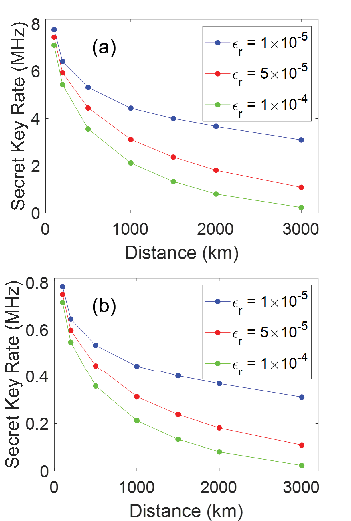}}
	\caption{secret key rates for communication distances between 100km and 3000km with different photon error probabilities: (a) $t_E$ = 10ns and $t_{CZ}$ = 10ns (b) $t_E$ = 100ns and $t_{CZ}$ = 100ns}
	\label{Fig4}
\end{figure}

In Fig. \ref{Fig4}, the number of quantum emitters is fixed to be 100. Separations between neighboring repeater nodes are restricted to be above 1km. I estimated the maximum rates for communication distances between 100km and 3000km with different photon error probabilities.

\section{Fault-tolerant Repeater}
5-qubit error correction code has been proposed to integrate with photonic cluster state to correct both photon loss errors and operational errors for quantum repeaters \cite{wo2023resource}. Correction of operational errors require tremendous amount of resources. Multiple quantum emitters can be a much more efficient tool for prepare the photonic graph state. More sophisticated error correction code need to be applied to one-way quantum repeaters to make them robust against operational errors and environmental noises for long distance entanglement distributions. 

\section{Discussions \& Conclusions}
In order to achieve high fidelity of photon-spin CZ gate, photons should have much longer wave packet compared with cavity decay time \cite{duan2004scalable}. The multiplexed scheme of photonic tree cluster state generation don't need photon-spin gate, which can greatly reduce processing time, and also there is no need of feedback loops for the photons. Atom array integrated with nanophotonic chips are also good candidates for multiplexed photonic graph state generation with strong interactions between Rydberg atoms enables high fidelity of gate operations and integration with nanophotonic chips make it possible to control each atom individually \cite{ocola2024control, menon2023integrated}. The proposal in this article paves the way towards one-way quantum repeaters with high communication rates.  

\section{Acknowledgments}
 Y.L. acknowledge the support from Northwestern University.

\section{Disclosures}
The author declares no conflicts of interest.

\section{Data availability}
Data underlying the results presented in this paper are not publicly available at this time but may be obtained from the authors upon reasonable request.

\bibliography{sample}{}

\end{document}